\documentstyle[12pt,cite ]{article}
 \makeatletter
\def\@cite#1#2{\textsuperscript{[{#1\if@tempswa , #2\fi}]}}
\input epsf
\title{\Large{\bf{Some Calculations for Cold Fusion Superheavy Elements }}
\thanks{Supported by National Natural Science
Foundation of China (10275037) and Specialized Research Fund for
the Doctoral Program of Higher Education of China (20010055012)}}

\author{ \footnotesize X.\ \ H. Zhong,  \footnote {E-mail: zhongxianhui@mail.nankai.edu.cn}
\, \,L.\ \ Li,  \footnote {E-mail: lilei@nankai.edu.cn}  \,\, P.\
\ Z. Ning \footnote {E-mail: ningpz@nankai.edu.cn}
\\\footnotesize  \emph{Department of Physics, Nankai University,
 Tianjin 300071, P. R. China}}
\begin{document}
\maketitle

\begin{abstract}
The $Q$ value and optimal exciting energy of the hypothetical
superheavy nuclei in cold fusion reaction are calculated with
relativistic mean field model and semiemperical shell model mass
equation(SSME) and the validity of the two models is tested. The
fusion barriers are also calculated with two different models and
reasonable results are obtained. The calculations can give  useful
references for the experiments in the superheavy nuclei
synthesized in cold fusion reactions.


\end{abstract}

   The formation of compound nuclei by cold fusion reaction   is
one of the outstanding problem of low-energy nuclear reactions.
Such processes play a key role in the production of superheavy
elements(SHEs)\cite{1,2,3,4}. Experiments or intensive preparatory
work are in progress at the laboratories such as Dubna, Berkeley,
RIKEN and GANIL etc. Theoretical support of these very expensive
experiments is vital in the choice of fusing nuclei and their
collision energy, and for the estimation of the cross sections and
identification of evaporation residues. The quantum mechanical
fragmentation theory(QMFT) which was developed by Gupta et al. to
describe the synthesis of new and superheavy elements gives us a
method for selecting an optimum cold target-projectile
combination. Cold compound systems were considered to be formed
for all those target+projectile combinations that lie at the
bottom of the potential energy minima\cite{5,6,7}. On the basis of
the QMFT, recently, Gupta lists all the possible target+projectile
combinations, referring to minima in PES, in particular, the use
of radioactive nuclear beams and targets\cite{8}.

There are several important physical values such as $Q$ values,
the heights of the fusion barrier $B_{fu}$ and the optimal
excitation energies $E^{*}$ for the cold fusion reaction. The aims
of this paper are to find some proper methods to predict the
ground-state $Q$ values, the heights of the fusion barrier
$B_{fu}$ and the optimal excitation energies $E^{*}$ for the cold
fusion reaction. Then, following the work of Gupta et al. in
Ref.\cite{8} we calculate $Q$, $B_{fu}$  and $E^{*}$  for the
possible target+projectile combinations to give some significative
references for the future experiments.

R. Smolanczuk  proposes an explanation of the production of
superheavy nuclei in the cold fusion reactions\cite{3}. In the
model, based on simple analytical formulas, they assume that the
neutron is evaporated from the compound nucleus formed by quantal
tunnelling through the fusion barrier. In the present paper we
adopt this model.

According to the Ref.\cite{3}, for the sake of simplicity, the
fusion barrier is used as the Coulomb potential cut off at the
distance $R_{fu}$, which depends on the size and the electric
charge of the colliding nuclei. The fusion barrier reads
\begin{equation}
B_{fu} =\frac{Z_{T}Z_{P}e^{2}}{R_{fu}} .
\end{equation}
The detail explanations of the formula are listed in Ref.\cite{3}.
We can also calculate in another new method proposed from a
fitting procedure on GLDM data on 170 fusion reactions by
Moustabchir et al.\cite{9}
\begin{equation}
B_{fu}=-19.38+\frac{2.1388Z_{1}Z_{2}+59.427(A_{1}^{1/3}+A_{2}^{1/3})-27.07ln(\frac{Z_{1}Z_{2}}{A_{1}^{1/3}+A_{2}^{1/3}})}{(A_{1}^{1/3}+A_{2}^{1/3})(2.97-0.12ln(Z_{1}Z_{2}))}.
\end{equation}
Since the  predictions of the GLDM agree with the experimental
data and the new formula reproduce the GLDM data more precisely
than other formulas for light and medium systems\cite{9}, we
extrapolate  eq.(2) to heavy system and expect to get good
prediction as well.

  The $Q$ value  is calculated with the formula
\begin{equation}
Q=E(CN)-E(A_{1})-E(A_{2})
\end{equation}
$E(CN),E(A_{1})$ and $E(A_{2})$ are the ground state binding
energies of compound nucleus, target and projectile nucleus,
respectively. In the present calculation, measured binding
energies of the target nucleus and the projectiles are
used\cite{10}. The compound nucleus binding energy $E(CN)$ is
calculated by relativistic mean field theory (RMF)\cite{11,12,13}
or the semiemperical shell model mass equation(SSME)\cite{14}. In
the SSME the total nuclear energy in the ground state is written
as a sum of pairing, deformation and Coulomb energies:
\begin{equation}
E(N,Z)=E_{pair}(N,Z)+E_{def}(N,Z)+E_{Coul}(N,Z).
\end{equation}

The total excitation energy $E^{*}$ is calculated as a difference
$E^{*}=E-Q$ between the bombarding energy $E$ and the ground state
$Q$ value for the considered reaction. Normally, we only care
about the optimal bombarding energy $E_{opt}=E_{opt}^{*}+Q$, the
optimal exciting energy $E_{opt}^{*}$ reads\cite{15}
\begin{equation}
E_{opt}^{*}=E(EV,eq)-E(CN,eq)
\end{equation}
$E(EV,eq)$ is the binding energy of the compound nucleus after the
evaporation of two neutrons and $E(CN,eq)$ is the binding energy
of the compound nucleus at equilibrium point.

When we calculate the  optimal exciting energy $E_{opt}^{*}$, the
key point is how to calculate the binding energy of compound
nucleus and evaporation residue at equilibrium point as close as
to the fact. Some groups use the constrained RMF \cite{15}to
calculate the binding energy at equilibrium point, but the
calculation is very heavy and complicated. For the sake of
simplicity we consider the ground state binding energy as the
binding energy of equilibrium point. In the present paper we use
the considered deformation RMF theory to calculate the $Q$ values
and optimal exciting energies $E_{opt}^{*}$, we also list the
results calculated from the semiempirical shell model mass
equation(SSME)\cite{14} to compare the two models with each other.

We carry out RMF calculations with the force parameters
NL3\cite{16}. The inputs of pairing gaps are
$\Delta_{n}=\Delta_{p}=11.2/\sqrt{A}$ MeV. The number of bases is
chosen as $N_{f}$=12, $N_{b}$=20. This space is enough for the
calculations here. For the details of calculations please see
Ref.\cite{13}.

In this work, first, we calculate the $Q$ values and optimal
exciting energies $E_{opt}^{*}$ of the reactions given in
Ref.\cite{3} with RMF and SSME models and also calculate the
fusion barrier of them with eq.(1), (2) and compare our results
with Ref.\cite{3} to test the reliability of our calculations,
then, we carry out the calculations of the reactions listed in
table 1 of Ref.\cite{8}.

From table 1 we can see RMF model and SSME model  give  close $Q$
value to macroscopic-microscopic calculations in Ref.\cite{3}. The
$Q$ value of RMF model and SSME model is larger than the result in
Ref.\cite{3}$1.5\sim8.3$ MeV and $8.6\sim13.5$ MEV respectively.
Generally, the difference becomes larger with the increasing of
the atomic number of compound nucleus. We believe that  the
reasons of the large difference in the large atomic number region
are that neither RMF nor SSME model  used in the rich neutron
region is accurate and the compound nucleus maybe have more
complicate structure which is not considered at all with the
increasing of the neutron number. In order to further compare RMF
model with SSME model, the experimental binding energies of
$^{256}No$ and $^{258}Rf$ \cite{17}together with binding energies
calculated from RMF model and SSME model are listed in table 2.
Comparing the calculations with the experimental data, we can see
that the RMF model calculates the binding energy better than SSME
model, thus the RMF model should be more accurate than SSME model
in calculating  the $Q$ values. In summary, both RMF and SSME
model are useful to estimate the $Q$ value, however, developing
more powerful models to calculate $Q$ values accurately need more
experimental data.

The optimal exciting energies of RMF model, SSME model and
macroscopic-microscopic model in Ref.\cite{3} have the difference
$1\sim2$MeV except the RMF model give  obviously  larger value
19.3 MeV for $^{74}Ge+^{208}Pb$ and 16.8 MeV for
$^{84}Kr+^{208}Pb$ than that of SSME and macroscopic-microscopic
model in Ref.\cite{3}. The experimental data of optimal exciting
energies is also listed in table 1, the difference of calculations
with experimental data is about $1\sim4$ MeV, although the
calculations are not very accurate, the results are fairly good
because we neglect many effects of the compound nuclear binding
energy at the equilibrium point. It is a success to calculate the
optimal exciting energies conveniently by the simple models.

\begin{table}
\begin{center}
\caption{\footnotesize Calculated ground-state $Q$ value, the
height of the fusion barrier $B_{fu}$ and optimal excitation
energy $E^{*}_{opt}$. Q and $E^{*}_{opt}$ are the results of
Ref.\cite{3}, $Q_{1}$ and $E^{*}_{opt1}$ calculated from RMF,
$Q_{2}$ and $E^{*}_{opt2}$ calculated from SSME, $E^{*}_{exp}$ is
the measured value. $B_{fu}$ is the value calculated by eq.(1) of
Ref.\cite{3}, $B_{fu1}$ is the value calculated by eq.(2).}
\label{parameter} \vspace{0.1cm} \footnotesize
\begin{tabular}{|c|c|c|c|c|c|c|c|c|c|c|c|c|c|c|c|c|}\hline\hline
HI &ER&$Q$ & $Q_{1}$&$Q_{2}$ & $B_{fu}$&$B_{fu1}$&$E_{opt}^{*}$&$E_{opt1}^{*}$&$E_{opt2}^{*}$&$E_{exp}^{*}$\\
\hline
$^{48}Ca$ & $^{255}No$  &153.6  &156.8   &162.5  &178.7 &192.4&13.28&12.41&13.33&16.7\\
$^{50}Ti$ & $^{257}Rf$  &169.6  &171.2   &178.2  &199.4&210.4&14.29&13.53&14.3&15.5\\
$^{54}Cr$ & $^{261}Sg$  &187.1  &188.3   &195.9  &219.1&227.2&14.65&13.9&14.4&16.4\\
$^{58}Fe$ & $^{265}Hs$  &204.9  &206.4   &213.9  &238.5&243.9&14.33&14.0&14.5&13.2\\
$^{62}Ni$ & $^{269}110$ &223.2  &224.8   &231.9  &257.5&260.5&13.33&14.3&14.8&13.2\\
$^{64}Ni$ & $^{271}110$ &224.5  &227.5   &234.3  &256.9&259.4&13.70&13.6&14.1&11.7\\
$^{68}Zn$ & $^{275}112$ &242.0  &244.2   &251.2  &275.6&275.8&13.07&15.1&14.3&\\
$^{70}Zn$ & $^{277}112$ &243.7  &247.0   &253.2  &274.9&274.7&12.62&12.9&13.7&\\
$^{74}Ge$ & $^{281}114$ &262.1  &263.8   &271.2  &293.4&291.0&11.89&19.3&14.0&\\
$^{76}Ge$ & $^{283}114$ &263.9  &267.0   &273.8  &292.8&290.0&11.50&12.7&13.4&\\
$^{80}Se$ & $^{287}116$ &282.5  &287.2   &285.6  &311.0&306.2&11.78&14.3&13.8&\\
$^{82}Se$ & $^{289}116$ &284.1  &290.7   &295.2  &310.3&305.1&12.12&12.6&13.4&\\
$^{82}Kr$ & $^{289}118$ &298.9  &303.5   &308.3  &329.6&323.3&12.31&12.4&14.7&\\
$^{84}Kr$ & $^{291}118$ &301.8  &304.8   &311.8  &329.0&322.2&12.81&16.8&14.4&\\
$^{86}Kr$ & $^{293}118$ &304.4  &308.9   &314.9  &328.3&321.2&13.31&13.8&13.9&\\
$^{78}Ge$ & $^{285}114$ &264.8  &270.9   &275.7  &292.1&289.0&11.29&11.0&12.9&\\
$^{80}Ge$ & $^{287}114$ &264.9  &272.4   &276.8  &291.5&287.9&11.35&12.2&12.5&\\
$^{82}Ge$ & $^{289}114$ &264.2  &272.5   &277.7  &290.8&287.0&11.79&12.6&12.1&\\
$^{84}Se$ & $^{291}116$ &284.9  &291.1   &296.8  &309.7&304.1&12.59&14.0&12.9&\\
$^{86}Se$ & $^{293}116$ &282.3  &290.3   &294.9  &309.0&303.1&13.19&11.6&12.7&\\
$^{88}Kr$ & $^{295}118$ &303.1  &309.2   &314.1  &327.7&320.15&13.58&13.8&13.6&\\
$^{90}Kr$ & $^{297}118$ &301.1  &308.1   &312.0  &327.0&319.15&13.45&12.5&13.3&\\
$^{92}Kr$ & $^{299}118$ &298.2  &306.4   &308.7  &326.3&318.17&12.81&11.6&13.2&\\
 \hline

 \hline
 \hline
 \end{tabular}
 \end{center}
 \end{table}

\begin{table}
\begin{center}
\caption{\footnotesize Calculated  binding energy $E _{RMF}$ , $E
_{SSME}$ with RMF and SSME model respectively ,   $E _{exp}$ is
the experimental data from Ref.\cite{17}. } \label{parameter}
\vspace{0.1cm} \footnotesize
\begin{tabular}{|c|c|c|c|c|c|c}\hline
Nucleus &$E_{exp} $&$E_{RMF} $ & $E_{RMF}- E_{exp}$ & $E_{SSME} $&$E_{SSME} -E_{exp} $\\
\hline
$^{256}No$ &1898.6  &1895.7& -2.9& 1890.1&-8.5 \\
\hline
$^{258}Rf$ & 1904.6&1903.0& -1.6&  1896.1&-8.5\\
\hline
\end{tabular}
\end{center}
\end{table}

\begin{table}[h]
\begin{center} \caption{\footnotesize Calculated ground-state $Q$ value and the height
of the fusion barrier $B_{fu}$ . $Q_{1}$
 calculated from RMF, $Q_{2}$
calculated from SSME, $B_{fu1}$ is the value calculated by eq.(1)
of Ref.\cite{3}, $B_{fu2}$ is the value calculated by eq.(2).}
\label{parameter} \vspace{0.1cm}\footnotesize
\begin{tabular}{|c|c|c|c|c|c|c|c|c|c|c|c|}\hline\hline
pro.+tar. & $Q_{1}$&$Q_{2}$  & $B_{fu1}$&$B_{fu2}$&pro.+tar.&$Q_{1}$&$Q_{2}$ & $B_{fu1}$&$B_{fu2}$\\
\hline
$^{48}Ca+^{210}Po$  & 159.8  & 165.2 &196.7 &183.6  &  $^{78}Zn+^{208}Pb$  &251.6  &256.8 &270.8 &272.4\\
$^{52}Ti+^{206}Pb$  & 172.8  &178.2  &209.8 &199.1  &  $^{82}Ge+^{204}Hg$  &261.9  &267.2 &281.0 &303.1\\
$^{84}Se+^{174}Yb$  & 232.5  &237.9  &267.9 &267.3  &  $^{134}Te+^{152}Nd$ &324.4  &329.7 &336.1 &346.6\\
$^{88}Kr+^{170}Er$  & 239.4  &244.8  &275.0 &275.3  &  $^{48}Ca+^{236}Pu$  &173.2  &180.1 &214.8 &205.8\\
$^{124}Sn+^{134}Xe$ & 275.9  &281.3  &300.1 &304.1  &  $^{78}Ge+^{206}Pb$  &267.5  &274.5 &289.5 &292.5\\
$^{48}Ca+^{212}Rn$  & 160.7  &168.3  &200.9 &188.5  &  $^{84}Se+^{200}Hg$  &277.3  &284.3 &298.8 &303.2\\
$^{52}Ti+^{208}Po$  & 174.7  &182.3  &214.5 &204.3  &  $^{86}Kr+^{198}Pt$  &285.1  &292.0 &308.1 &313.5\\
$^{56}Cr+^{204}Pb$  & 188.2  &195.8  &227.1 &219.0  &  $^{136}Xe+^{148}Nd$ &335.7  &342.6 &349.9 &360.9\\
$^{86}Kr+^{174}Yb$  & 248.0  &255.6  &282.8 &283.8  &  $^{50}Ca+^{238}Pu$  &173.3  &179.1 &213.4 &204.9\\
$^{122}Sn+^{138}Ba$ & 286.0  &293.6  &310.6 &315.6  &  $^{80}Ge+^{208}Pb$  &271.1  &276.8 &287.9 &291.5\\
$^{50}Ca+^{220}Ra$  & 158.0  &164.4  &198.5 &186.9  &  $^{84}Se+^{204}Hg$  &280.6  &287.3 &314.8 &302.5\\
$^{62}Fe+^{208}Pb$  & 209.4  &215.8  &241.8 &237.2  &  $^{134}Te+^{154}Sm$ &334.8  &340.6 &346.7 &358.1\\
$^{68}Ni+^{202}Hg$  & 219.6  &225.9  &252.5 &249.7  &  $^{50}Ca+^{240}Pu$  &174.3  &179.2 &213.0 &204.7\\
$^{84}Se+^{186}W$   & 247.2  &253.6  &279.9 &281.5  &  $^{82}Ge+^{208}Pb$  &272.0  &276.9 &287.0 &290.9\\
$^{134}Xe+^{136}Xe$ & 303.3  &309.6  &319.6 &326.8  &  $^{84}Se+^{206}Hg$  &281.7  &286.6 &297.2 &302.2\\
$^{50}Ca+^{206}Pb$  & 157.9  &163.9  &202.7 &191.7  &  $^{134}Te+^{156}Sm$ &336.6  &341.5 &345.0 &357.6\\
$^{52}Ti+^{222}Ra$  & 171.5  &177.5  &217.1 &208.4  &  $^{50}Ca+^{244}Pu$  &175.3  &178.5 &212.3 &204.2\\
$^{84}Se+^{188}W$   & 247.3  &253.3  &279.4 &281.2  &  $^{132}Sn+^{162}Gd$ &336.4  &339.6 &342.0 &353.9\\
$^{94}Sr+^{178}Yb$  & 261.2  &267.2  &293.7 &297.6  &  $^{48}Ca+^{242}Cm$  &179.9  &185.3 &218.2 &210.1\\
$^{136}Xe+^{136}Xe$ & 305.5  &311.5  &318.8 &326.3  &  $^{50}Ca+^{240}Cm$  &178.4  &183.7 &217.2 &209.6\\
$^{48}Ca+^{222}Th$  & 163.3  &170.5  &208.3 &197.4  &  $^{84}Se+^{206}Pb$  &290.3  &295.6 &304.7 &310.1\\
$^{66}Ni+^{204}Pb$  & 227.4  &234.6  &259.3 &256.9  &  $^{136}Xe+^{154}Sm$ &338.4  &343.7 &359.4 &371.7\\
$^{86}Kr+^{184}W$   & 265.2  &272.4  &296.1 &299.1  &  $^{48}Ca+^{246}Cf$  &184.0  &190.0 &222.0 &209.6\\
$^{96}Zr+^{174}Yb$  & 278.7  &285.8  &309.6 &314.5  &  $^{86}Kr+^{208}Pb$  &308.9  &314.9 &321.2 &328.4\\
$^{132}Xe+^{138}Ba$ & 313.8  &321.0  &331.6 &339.4  &  $^{136}Xe+^{158}Gd$ &361.0  &389.0 &369.6 &383.0\\
$^{48}Ca+^{230}U$   & 167.8  &174.4  &211.4 &201.5  &  $^{50}Ca+^{252}Fm$  &188.6  &190.5 &224.3 &218.0\\
$^{72}Zn+^{206}Pb$  & 247.1  &253.8  &273.7 &274.7  &  $^{94}Sr+^{208}Pb$  &326.5  &328.4 &335.1 &344.8\\
$^{78}Ge+^{200}Hg$  & 256.6  &263.2  &284.0 &285.9  & $^{136}Xe+^{166}Dy$  &374.9  &376.8 &378.3 &393.2\\
$^{84}Se+^{194}Pt$  & 265.9  &272.6  &292.9 &296.3  &&&&&\\
$^{138}Ba+^{140}Ba$ & 326.7  &333.4  &340.8 &350.4  &&&&&\\
$^{50}Ca+^{236}U$   & 169.0  &174.2  &209.2 &200.2  &&&&&\\

\hline \hline
\end{tabular}
 \end{center}
 \end{table}

\begin{table}
\begin{center}
\caption{ \footnotesize Calculated  optimal excitation energy
$E^{*}_{opt}$ , $E^{*}_{opt1}$ calculated from RMF, $Q_{2}$ and
$E^{*}_{opt2}$ calculated from SSME.} \label{parameter}
\vspace{0.1cm} \footnotesize
\begin{tabular}{|c|c|c|c|c|c|c|c|c|c|c|c|c|c|c|c|c|}\hline\hline
Nucleus &$E_{opt1}^{*}$&$E_{opt2}^{*}$ &Nucleus&$E_{opt1}^{*}$&$E_{opt2}^{*}$&Nucleus&$E_{opt1}^{*}$&$E_{opt2}^{*}$\\
\hline
$^{258}Rf$ &12.1  &14.3&$^{270}110$  &14.3&14.8&$^{288}114$&13.5&12.5 \\
$^{260}Sg$ & 13.9 &15.1&$^{278}112$ &13.4&13.7&$^{290}114$&11.2&12.1 \\
$^{270}Hs$ & 13.1 &13.2&$^{286}112$ &10.6&11.6&$^{294}114$&13.3&11.6 \\
$^{272}Hs$ & 12.2 &12.6&$^{284}114$ &12.9&13.4&$^{290}116$&12.9&13.4 \\
           &  &        &$^{294}118$ &14.8&13.9&$^{302}120$&12.3&14.1 \\
 \hline

 \hline
 \hline
 \end{tabular}
 \end {center}
 \end{table}
Finally, let's see the fusion barrier in table 1, the barrier
calculated by eq.(1) is lower than that calculated by eq.(2) till
to atomic number of compound nucleus $A=276$, the maximum
difference is 13.7MeV for $^{48}Ca+^{208}Pb$ and the minimum
difference is 0.2MeV  for $^{68}Zn+^{208}Pb$. When $A\ge276$, the
barrier calculated by eq.(1) is higher than that calculated by
eq.(2), the difference increases from 0.2MeV to 8MeV with the
increasing of atomic number. As a whole both eq.(1) and eq.(2) can
give close values of fusion barrier and it indicates that eq.(2)
can be extrapolated to heavy system, though there are differences
in the calculating. In the following work, we calculate the fusion
barrier with eq.(1) and (2) and give the result in table 3.

  We calculate the $Q$ value and optimal
exciting energy $E_{opt}^{*}$ with RMF and SSME models and fusion
barrier $B_{fu}$ with eq.(1),(2) of the reactions given in
Ref.\cite{8} and the results listed in table 2, 3.

 From table 2 we can see the $Q$ value of RMF models is less than
 SSME model $5\sim8$ MeV. Some of the fusion barriers calculated by eq.(1)
 and (2) are very close to each other such as $^{84}Se+^{174}Yb$ and
 $^{88}Kr+^{170}Er$, the variety is within 1 MeV,
 most of them have the difference
 about 10 MeV, For $^{82}Ge+^{204}Hg$, there is large
 difference 20.2 MeV.
 In table 3, the optimal exciting energies of RMF and SSME
 models are given, the calculated results are about 11 to 15 MeV,
 the differences between  the two models are about 1 MeV except there
 are larger differences about 2 MeV for  $^{258}Rf$ and $^{302}120$ respectively.

 In summary, we use different models to calculate the ground-state
 $Q$ value, optimal exciting energy and fusion barriers of
 superheavy nuclei listed in Ref.\cite{3} in cold fusion reaction.
 These models can reproduce the results of Ref.\cite{3}, thus, we
 also use these models  to predict the ground-state
 $Q$ value, optimal exciting energies and fusion barriers
 corresponding to the reactions suggested by Gupta\cite{8} in the future
 experiments. The theoretical results can be used for a guide of future experiments
 of superheavy nuclei synthesized in cold fusion reactions.

\end{document}